\documentstyle[12pt,german]{article}
\renewcommand{\baselinestretch}{1.5}
\begin{document}
\begin{quotation}
\noindent
\hfill MPI--PhT/96--32\\
\begin{center}
{\large The problem of Mass and
Mass Generation\renewcommand{\thefootnote}
{\fnsymbol{footnote}}\footnote[1]{Supported by EC Projects 
Athen contract--no. SCI--CT 91--0729, Marseille contract--no. CHRX--CT--0579
and GIF Israel contract--no. I 0304--120.07/93}\renewcommand{\thefootnote}
{\fnsymbol{footnote}}\footnote[2]{Invited talk given at the Warsaw Meeting
``Physics at the turn of the century'' from 29 -- 30 January 1996}}
\end{center}
\bigskip
\begin{center}
{\large Harald Fritzsch}\\
\end{center}
\begin{center}
{\it Ludwig--Maximilians--Universit"at, Sektion Physik,\\
Theresienstra"se 37, D--80333 M"unchen\\
\bigskip
Max--Planck--Institut f"ur Physik, Werner--Heisenberg--Institut,\\
F"ohringer Ring 6, D-80805 M"unchen\\
E--Mail: bm@hep.physik.uni--muenchen.de}
\end{center}
\newpage
\noindent
{\large \normalsize Towards the end of
the last century the electron was
discovered. In retrospect this discovery marked
the beginning of a remarkable development, which eventually led to the
emergence of the ``Standard Model of Fundamental Particles and Forces''
in the 70ies. According to the letter all the visible matter in the
universe is composed of fundamental objects of two different categories
-- leptons (among them the electron) and quarks. The latter do not exist
as free particles, but are bound among each other to form the protons and
neutrons, the building blocks of the atomic nuclei.\\
\\
The dynamics of matter in our universe can be traced back to the action of
four types of fundamental forces: the strong forces among the quarks,
the electromagnetic forces among charged particles, the weak interactions
responsible for the phenomenon of radioactivity, and gravity. With the
exception of the strong forces all other interactions were known, at least
through their indirect effects, at the turn of the last century. The first
effects
of the strong nuclear force were found in 1919 by
Rutherford and his collaborators by studying the interactions of
$\alpha$--particles and nitrogen nuclei. Finally in 1932 the neutron was found
by Chadwick and soon thereafter identified as a second building block
of the atomic nucleus besides the proton.\\
\\
The ``Standard Model''
constitutes a consistent theory of the fundamental forces, based
on quantum field theory and the concepts of non--Abelan gauge theories.
Since about 1980 the quantitative predictions of the ``Standard Model'' have
been subjected to severe experimental tests, with the result that no
departures from the theoretical expectations have been found. The
``Standard Model'' provides us not only with a fairly good description
of the fundamental particles and forces, but gives an excellent picture
of reality. Not a single confirmed result from the particle physics
experiments are in conflict with it.\\
\\
With the help of the Large
Electron Positron collider (LEP) at CERN one was able in the recent years
to test the predictions of the ``Standard Model'', especially its
electroweak sector, with a high order of precision. Furthermore the study
of the collisions at LEP has helped to unify particle physics, astrophysics
and cosmology to a coherent picture of the cosmic evolution. The collisions
at LEP recreate the conditions which were present in the universe about
$10^{-10}$ seconds after the Big Bang.\\
\\
According to the ``Standard Model'' there exist two different categories
of fundamental particles: the matter particles (quark, leptons) carrying
spin 1/2
and the force particles (photons, $W$, $Z$, gluons) carrying spin 1.
The latter are gauge bosons, and their interactions with the matter particles
and with themselves are dictated by the principles of non--Abelean gauge
symmetry. The symmetry group is given by the direct product of three simple
groups:
$SU(3)_c \times SU(2)_w \times U(1)$ (c: color, w: weak).\\
The gauge bosons are given in the following table:\\
\begin{center}
\hspace*{1cm}Gauge Bosons (spin 1)\\
\end{center}
\begin{center}
\begin{tabular}{cccc}
\setlength{\unitlength}{1cm} 
\begin{picture}(2,1)
\put(1,0){\line(1,0){1}}
\put(1,0){\line(0,1){0.3}}
\put(1,0.3){\line(1,0){1}}
\put(2,0.3){\line(0,-1){0.3}}
\end{picture}
         & Mass     & Electric Charge & Color\\
\hspace*{1.0cm} $\gamma$ &  0       &     0           & 0\\
\hspace*{1.0cm} $W^-$    & 80.2 GeV &    $-$1           & 0\\
\hspace*{1.0cm} $W^+$    & 80.2 GeV &    $+$1           & 0\\
\hspace*{1.0cm} $Z^ø$    & 91.2 GeV &     0           & 0\\
\hspace*{1.0cm} g        &   0      &     0           & 8\\
\end{tabular}
\end{center}
\newpage
\noindent
The visible matter in the universe is composed of the elements of the first
lepton--quark family:\\
\[ \left( \begin{array}{rrrrr}
        \nu_e & \vdots &  u_{r} & u_g & u_b\\ 
        e^-   & \vdots &  d_{r} & d_g & d_b
\end{array} \right) \]
($r$, g, b: color index).\\
\\
The fact that the electric charges of these eight objects add to zero
indicates that the two leptons and six quarks are related to each other
in a way which cannot be described within the ``Standard Model'', but is
subject to the yet hypothetical physics beyond the ``Standard Model'',
perhaps described by a grand unification of all interactions.\\
\\
All known particles can be described in terms of three
lepton--quark--families:\\
\\
\[ \begin{array}{ccc}
\left( \begin{array}{lll}
          \nu_e & \vdots & u\\
          e^-   & \vdots & d
          \end{array} \right) & 
          \left( \begin{array}{lll} 
          \nu_{\mu} & \vdots & c\\
          \mu^- & \vdots & s
          \end{array} \right) &
          \left( \begin{array}{lll}
          \nu_{\tau} & \vdots & t\\
          \tau^- & \vdots & b \end{array} \right) \, .\\
\\
I & II & III\\ \\
\end{array} \]
\noindent
It should be stressed that several algebraic properties of the
``Standard Model'' cannot be deduced from the underlying symmetry, e.\ g.\
the charge and spin assignments of the leptons and quarks, or the number
of families. It is remarkable that the number ``three'' plays a three--fold
significant role:
\begin{enumerate}
\item[a)] Quarks come in three colors, hence the nucleons consist of
three quarks.
\item[b)] There exist three families of leptons and quarks.
\item[c)] There are three different gauge interactions, based on the
three different gauge groups (gravity is not considered here).
\end{enumerate}
The question arises whether there are connections between three different
``threenesses'' of the Standard Model. For example, the number of families
and colors could be related by an underlying, yet unknown symmetry principle.\\
\\
The most unsatisfactory feature of the ``Standard Model'' is the fairly
large number of free parameters which need to be adjusted. First of all,
the values of the three gauge coupling constants $g_1, g_2, g_3$ have to
be taken from experiment. All other parameters are related to the masses
of the fermions or gauge bosons. They are: the mass of the $W$--boson,
the masses of the three charged leptons $m_e, m_{\mu}, m_{\tau}$ (the
neutrinos will be kept massless in our discussion), the masses of the six
quark flavors $m_u, m_d, m_c, m_s, m_t, m_b$ and the four mixing angles of the
weak interaction sector $\Theta_{sd}, \Theta_{sb}, \Theta_{bd}$ and $\delta $ 
($\Theta_{sd}$ denotes the angle, which describes primarily the s -- d
mixing etc., $\delta $: phase angle). Thus 17 parameters are needed to
describe the observed particle physics phenomena.\\
\\
We point out that 13 free parameters of the ``Standard Model'' are associated
with the masses of the matter fermions (nine masses, four mixing parameters,
which arise due to a mismatch between the mass matrices of the $u$--type
quarks and the $d$--type quarks). Thus a deeper insight into the dynamics of
the mass generation for the leptons and quarks is needed in order to reduce
the number of free parameters -- an insight which would definitely carry
us beyond the physics of the ``Standard Model''. Certainly the problem
of mass and mass generation are on the top of the priority list of problems
in fundamental physics for the immediate future.\\
\\
In all physics phenomena the different interactions and mass scales enter in
a variety of ways, thus producing the multitude of phenomena we observe.
Before discussing the problem of the lepton and quark masses, let me stress
that the problem of the nucleon mass (and therefore of the masses of the
atomic nuclei, which represent more than 99\% of the mass of the visible
matter in the universe) has found an interesting solution within the
framework of QCD. Due to the renormalization effects the QCD gauge coupling
constant $\alpha_s$ depends on the scale at which the interaction is
studied. It decreases logarithmically at high energies. At a scale $\mu $
its decrease is given by $\alpha_s(\mu^2) = \rm const. / \rm ln (\mu^2 / \Lambda^2)$,
where $\Lambda $ is a scale parameter, which serves as the fundamental mass
scale of the theory, i.\ e.\ a mass scale which fixes all other mass scales
in strong interaction physics (the effects of the quark masses are neglected).
Phenomenologically $\Lambda $ is about $150 \ldots 200$ MeV, i.e.
$\Lambda^{-1}$ corresponds to the typical extension of a hadron.\\
\\
Using
powerful computers and sophisticated nonperturbative methods one has been
able to calculate the masses of the lowest--lying hadrons (nucleon,
$\Delta$--resources, $\rho $--mesons $\ldots $) in terms of $\Lambda $,
with impressive results. Escpecially the mass ratios
$(m_{\Delta } / m_p), m_{\rho }/ m_p)$ etc.\ can be calculated with
high precision. 
Thus we can say that the problem of the nuclear masses,
especially of the proton mass, has found a solution. The mass
of a proton (about 940 MeV) is nothing but the field energy of the quarks
and gluons, which are confined within a radius of the order of $10^{-13}$ cm
(of order $\Lambda^{-1}$). Therefore, a direct link exists between the nucleon
mass and the size of the nucleon. Both have to be of the same order of
magnitude -- more specifically the nucleon mass is expected to be of the
order of $3 \cdot \Lambda$, where 3 denotes the number of the constituent
quarks. Furthermore the nucleon mass is a truly nonperturbative phenomenon,
directly related to the confinement aspect of QCD. The QCD gauge interaction
itself creates its own mass scale -- the nuclear mass is generated dynamically.\\
\\
This phenomenon of dynamical mass generation is not directly related to the
quark substructure of the hadrons, but rather to the gluonic degrees of
freedom. This can be seen by studying the mass spectrum of ``pure QCD'',
i.\ e.\ QCD without quarks. This theory of eight interacting gluons displays
at low energies a discrete mass spectrum, which starts at the mass of the
lowest lying glue meson. Thus unlike ``pure QED'' the theory displays
a mass gap which is generated by nonperturbative effects.\\
\\
The theory of the nucleon mass described above is remarkable in the sense
that the mass of an elementary particle can in principle be calculated.
Note that the mass is directly related to the field energy density inside
the nucleon.
It would be of high interest to know whether the masses of the leptons, the
$W$--, $Z$--particles and the quarks are due to a similar mechanism, or are
generated by a qualitatively different mechanism.\\
\\
In the standard electroweak theory these masses are due to a spontaneous
breaking of the electroweak gauge symmetry caused by an elementary scalar
field $\varphi $.
The order parameter of the symmetry breaking is given by the vacuum expectation
value $v $ of the field $\varphi $ which in turn is related to the
Fermi constant $G$ and the $W$--mass:\\
\begin{displaymath}
\frac{G}{\sqrt{2}} = \frac{g_{w^2}}{8 M_{w^2}} = {1}{2 v^2}\\
\end{displaymath}
\bigskip
($g_w$: gauge coupling constant).\\
\\
The main consequence of this mechanism is a relation between the $Z$--mass
and the $W$--mass in terms of the electroweak mixing angle $\Theta_w$:
$M_z = M_w / \cos \Theta_w$. Since the mixing angle $\Theta_w$ can be
determined independently by studying the neutral current interaction of
leptons and quarks, this mass relation is a nontrivial constraint, in
excellent agreement with the experimental data.\\
\\
The success of the electroweak mass relation does not necessarily imply
that the
mechanism of the spontaneous symmetry breaking is realized in the real
world, but it implies that an alternative mechanism must lead to the same
mass relation. This is the case, for example, in technicolor models in which
the scalar field $\varphi $ is replaced by a field composed of new fermions
which are tightly bound by the new technicolor interaction.\\
\\
If the standard electroweak model is correct, it implies the existence of
a particle, the ``Higgs'' particle, whose couplings are given by the observed
particle masses. The mass of this particle is unknown, but it can hardly be
larger than about 1000 GeV. The LEP experiments exclude the mass region
lower than about 60 GeV.\\
\\
Certainly the most interesting question is the one about the origin
of the lepton-- and quark masses. The spectrum extends over five orders
of magnitude, starting with the electron (neutrino masses are not considered
here), and ending with the $t$--quark with a mass of about 180 GeV. Thirteen
free parameters are needed to describe the properties of the lepton and quark
mass spectrum: the three lepton masses, the six quark masses, and the four
parameters describing the mixing of the quark flavors. Unlike the masses of
the leptons, the quark masses cannot be determined directly, but have to be
inferred from the properties of the hadronic spectrum. Furthermore they are
scale dependent, i.\ e.\ they vary logarithmically, if the corresponding
renormalization point is shifted. In lowest order this change is given by
$m_q(\mu) = m_q(\mu_0) (1 - \frac{\alpha_s (\mu)}{\pi}$ ln $\frac{\mu^2}
{\mu{_0}{^2}})$, where $\alpha_s(\mu)$ is the QCD coupling constant. A
suitable renormalization point for the quark masses is the mass of the
$Z$--boson, which is known with a high precision:
$M_z = 91.1884 \pm 0.0022 $ GeV.\\
One finds:
\[
\begin{array}{rrrrrrrrrrrrr}
m_u(M_z) & = & 3.4 & \pm & 0.6 & MeV, & m_d(M_z) & = & 6.3 & \pm & 0.9 & MeV\\
m_c(M_z) & = & 880.0 & \pm & 48.0 & MeV,  & m_s(M_z) & = & 118.0 & \pm & 17.0 & MeV\\
m_t(M_z) & = & 17.0  & \pm & 12.0 & GeV,   & m_b(M_z) & = & 3.31 & \pm & 0.11 & GeV
\end{array}
\]
\\
A closer inspection of the mass spectrum tells us:
\begin{enumerate}
\item[a)] The mass spectra of the three flavor channels (charged leptons,
$u$--type quarks, $d$--type quarks are almost entirely dominated by the mass
of the member of the third generation.
\item[b)] The relative importance of the second generation decreases as we
proceed upwards in the charge ($\mu \rightarrow s \rightarrow c$). In the
lepton case the muon contributes about 5.6\% to the sum of the masses, while
in the charge (--1/3)-- channel the $s$--quark contributes only 3.2\%, and in
the charge (+2/3)-- channel the $c$--quark contributes only 0.5\%.
\item[c)] The relative importance of the masses of the members of the first
generation is essentially zero.
\item[d)] The entire mass spectrum of the leptons and quarks is dominated
fairly well by the $t$--quark alone. For example, in the case $m_t = 100$ GeV
the $t$--quark contributes 97.5\% to the sum of all fermion masses. All
other quarks, mostly the $b$--quark, contribute only 2.5\%.\\
\end{enumerate}
The spectrum exhibits clearly a hierarchical pattern: The masses of a
particular generation of leptons or quarks are small compared to the masses
of the following generation, if there is any, and large compared to the
previous one if there is any. Furthermore another hierarchical pattern
emerges if we consider the weak interaction mixing parameters. The mixing
matrix, if written in terms of quark mass eigenstates, is not far from the
diagonal matrix (no mixing). The mixing angles are typically rather small;
the Cabibbo angle being the largest of all, is about 13$^{\circ}$,
while $\Theta_{sb}$ is about 2.2$^{\circ}$ and $\Theta_{bd}$ about
0.2$^{\circ }$.\\
\\
What kind of symmetry could one discuss in view of the observed lepton--quark
mass spectrum? The observed two different hierarchies suggest that we are
very close to a limit, which I like to call the ``rank 1''--limit, in which
both the $u$--type and $d$--type mass matrix can be diagonalized at the same
time and in which they both take the diagonal form (0, 0, 1), multiplied
by $m_t$ or $m_b$ respectively. Thus the masses of the first two generations
vanish (the mass matrix has rank one), likewise all mixing angles.\\
\\
Of course,
it depends on yet unknown details of the mass generation mechanism whether
such a limit can be achieved in a consistent way. We simply assume that this
is the case. In this limit there exists a mass gap: The third generation is
split from the massless first two generations. Obviously nature is not far
away from this limit, and therefore one is invited to speculate about the
dynamical origin of such a situation. A mass matrix proportional to the matrix
\[ \left[ \begin{array}{ccc}
       0 & 0 & 0\\ 
       0 & 0 & 0\\
       0 & 0 & 1
\end{array} \right] \]
\\
\\
can always be obtained from another matrix, namely the one in which all
elements are equal:\\
\[ \left[ \begin{array}{ccc}
       1 & 1 & 1\\
       1 & 1 & 1\\
       1 & 1 & 1
\end{array} \right] \]
\\
\\
by a suitable unitary transformation. Matrices of this type, which might be
called ``democratic mass matrices'' have been considered recently by a
number of authors.\\
They can be used as a starting point to
construct the full mass matrices of the quarks, including the weak
interaction mixing terms. We note that such a matrix plays
an important role in other fields of physics, where mass gap phenomena
are observed:\\
\begin{enumerate}
\item[a)] In the BCS theory of superconductivity the energy gap is related
to a ``democratic'' matrix in the Hilbert space of the Cooper pairs.
\item[b)] The pairing force in nuclear physics which is introduced in order
to explain large mass gaps in nuclear energy levels has the property that
the associated Hamiltonian in the space of nucleon pairs has equal matrix
elements, i.\ e.\ it has a structure of the type given above.
\item[c)] The mass pattern of the pseudoscalar mesons in QCD in the chiral
limit $m_u = m_d = 0$. In this limit the $\pi^{\circ}$ and the $\eta $ are
massless Goldstone bosons, while the $\eta' $ acquires a mass due to the
gluon anomaly.\\
\end{enumerate}
Once we write
the mass matrices of the leptons and quarks in their
``democratic'' form, it is obvious that there exists a symmetry, namely
the symmetry $S_3$ of permutations among the three different flavors.
This symmetry suggests that one should consider the eigenstates of the
quarks and leptons in \underline{this} basis as the fundamental dynamical
entities. Let us denote them as $(l_1, l_2, l_3)$ and $(q_1, q_2, q_3)$
respectively.\\
The heaviest lepton and quark, i.\ e.\ the $\tau $--lepton, the $t$ and $b$
quarks, would be \underline{coherent} states of the type:\\
\begin{displaymath}
\tau = \frac{1}{\sqrt{3}} (l_1 + l_2 + l_3) \hspace*{1.0cm} etc. \\
\end{displaymath}}
\renewcommand{\baselinestretch}{1.5}{\large \normalsize In view of the
scarce information we have at present about the internal
dynamics of the leptons and quarks we do not know, whether this description
of the fermions in terms of coherent states is more than a specific
mathematical representation. In a composite model, for example, the fermion
states $f_1, f_2, f_3$ would be those states which are ``pure'' in a
dynamical sense, e.\ g.\ they have simple unmixed wave functions.\\
\\
We remind the reader that also in the case of superconductivity and of the
nuclear pairing force the mass eigenstates are coherent superpositions of
``physical'' states which are described by simple wave functions (e.\ g.\
the Cooper pairs in superconductivity).\\
\\
Within our approach we see a solution to a problem, which has plagued many
models of the physics beyond the standard model, the problem of the near
masslessness of the first and to some extent also of the second generation.
In the coherent state basis this is easily understood. For example, the
electron state $e = 1 \sqrt{2} (l_1 - l_2)$ is nearly massless, since there
is a nearly complete cancellation of the $l_1$--  and $l_2$--mass terms, as
a consequence of the rank one structure of the dominant lepton--quark
mass term.\\
\\
Although this is not the place to discuss the details of the mass generation
for the first two generations, it is important to note that in various
dynamical schemes, in particular in one based on a composite structure of
the leptons and quarks, the introduction of these small masses leads to
slight breakings of the flavor conservation, especially in reactions
associated with large momentum transfers. In particular the reaction
$e + p \rightarrow \tau + X$, a reaction, which might be found at the
HERA machine in Hamburg,
is of interest. Moreover rare decays like $t \rightarrow Z + c$ can occur.\\
\\
In this talk I have described a number of ideas which one might consider after
looking at the pattern of masses exhibited in the lepton--quark mass spectrum.
I have emphasized the role of symmetries in the space of the generations of
the quarks in providing relations between the various mass eigenvalues and
the mixing angles. An approach to the flavor problem and to the hierarchical
mass spectrum of the leptons and quarks, based on the introduction of coherent
states, was discussed. It was argued that the mass generation for the third
lepton--quark generation is nothing but a gap phenomenon and is rather similar
to the mass generation for the pseudoscalar mesons in QCD. Thus the third
lepton--quark generation is somewhat distinct from the other ones. The same
mechanism which leads to the mass generation causes the appearance of
flavor changing effects; only in the absence of the lepton and quark masses
of the first and second generation the various quark and lepton flavors
are conserved.\\
\\
If our interpretation of the mass gap seen in the lepton--quark spectrum is
correct, it would mean that all mass gap phenomena seen in physics
-- superconductivity, nuclear pairing forces, QCD mass gap, lepton--quark
mas spectrum -- are due to an analogous underlying dynamical mechanism.
The exploration of further details of this mechanism could lead soon to a
deeper understanding of the physics beyond the standard model.}\\
\end{quotation}
\end{document}